\documentclass[12pt]{article}
\begin{document}
\begin{center}
{\large EFFECTIVE FIELD THEORY FOR NONCOMMUTATIVE SPACETIME:\\
A TOY MODEL }\\
\vskip 2cm
Subir Ghosh\\
\vskip 1cm
Physics and Applied Mathematics Unit,\\
Indian Statistical Institute,\\
203 B. T. Road, Calcutta 700108, \\
India.
\end{center}
\vskip 3cm
{\bf Abstract:}\\
A novel geometric model of a noncommutative plane has been
constructed. We demonstrate that it can be construed as a toy
model for describing and explaining the basic features of physics
in a noncommutative spacetime from  a field theory perspective.
The noncommutativity is induced internally through constraints and
does not require external interactions. We show that the
noncommutative space-time is to be interpreted as having an {\it
internal} angular momentum throughout. Subsequently, the
elementary excitations - {\it i.e.} point particles - living on
this plane are endowed with a {\it spin}. This is explicitly
demonstrated for the zero-momentum Fourier mode. The study of
these excitations reveals in a natural way various {\it stringy}
signatures of a noncommutative quantum theory, such as dipolar
nature of the basic excitations \cite{jab} and momentum dependent
shifts in the interaction point \cite{big}. The observation
\cite{sw} that noncommutative and ordinary field theories are
alternative descriptions of the same underlying theory, is
corroborated here by showing that they are gauge equivalent.

Also, treating the present model as an explicit example, we show
that,  even classically, in the presence of additional
constraints, (besides the usual ones due to reparameterization
invariances), the equivalence between Nambu-Goto and Polyakov
formulations is subtle.

\vskip 3cm \noindent Key Words: Noncommutative spacetime,
Constraints, Reparameterization invariant theory, Spinning
particle.

\newpage
Non-Commutativity in Quantum Field Theory (QFT) has a long history.
It was originally postulated by Snyder \cite{sny} as a means of a (Lorentz invariant)
regularization to cure the short distance singularities. In a different scenario,
Connes \cite{con} introduced Non-Commutative (NC) geometry by extending the standard
differentiable manifold to a mixed one with an additional discrete NC manifold, where
the Higgs field, (and subsequently the Higgs potential), appear as part of the gauge
field structure. For various reasons the above formulation did not gain much popularity.

With the advent of D-brane solutions \cite{pol} in open string
theory, the role of noncommutativity in space-time \cite{rev} has
gained importance since it reflects the low energy stringy
behavior in a (constant) background field. The effective NC gauge
theory is more tractable than the original string theory as the
former is (computationally) very close to field theory in ordinary
space-time. The NC is induced in the D-branes when the open string
endpoints are on the branes, in the presence of a two-form
background field, via a qualitative change in the boundary
conditions \cite{chu,hrt}.

A very interesting feature of NC theory is that in the charged
sector, the basic excitations act as {\it dipoles} \cite{jab} in
$U(1)$ gauge interactions. Another peculiarity is that a
consistent quantization requires a momentum dependent {\it shift}
in the interaction point \cite{big}, leading to a specific kind of
non-locality. A further crucial observation of \cite {sw} is that
ordinary and NC gauge theories are {\it alternative} descriptions,
(dictated by the choice of regularization), of the same quantum
theory.

The goal of this Letter is to remove some of the mysteries
surrounding NC theory and to establish the noncommutativity from
fundamental principles. In the present work we demonstrate that
the above mentioned alien features of an NC QFT can be
accommodated in a natural way, staying within the realm of
conventional QFT on a differentiable manifold. Thus their origin
can be studied in a unified manner at a deeper level. \footnote
{Generally, one considers the effective theory as an NC field
theory on the brane and simply postulates the fundamental NC brane
coordinate relation $[X^{\mu},X^{\nu}]=i\theta^{\mu\nu} $ and
proceeds. Also the properties \cite{jab,big} are stated without
explanation.} This requires the construction of a noncommutative
space-time hypersurface on which a field theory can be studied
directly. Essentially the present work is a toy model of such a surface.

Here we have initiated the study of  a relativistic,
reparameterization invariant theory of such a noncommutative
{\it{surface}}, both in the Nambu-Goto and Polyakov formalism. We
stress that, on this plane, the string induced effects of
noncommutativity mentioned above, are generated and can be
interpreted  in a natural way. The NC is induced internally
through constraints and does not require an external interaction.
Obviously, it can be reduced to conventional NC theory for the
sake of comparison. On this NC surface, we show that "point"
particles will behave like dipoles \cite{jab} in electromagnetic
interaction. Occurrence of the non-locality \cite{big}, via the
momentum dependent phase shift in the interaction vertex, is
explained naturally. Also, we demonstrate that in the Polyakov
formulation, the NC is induced, depending on our choice of gauge.
This is reminiscent of the equivalence between the NC and ordinary
gauge theories \cite{sw}. In the present setup, the celebrated
Seiberg-Witten map \cite{sw} might possibly be interpreted as a
gauge transformation.

The above has been achieved by the extension of the target space
$X^{\mu}(\sigma ,\tau )$ to$(X^{\mu}(\sigma ,\tau ),\\
N^{\nu}(\sigma ,\tau ))$ where $N^{\mu}N_{\mu}=1$. Hence the
extended manifold is a bundle of unit spheres over the flat
$X^{\mu}(\sigma ,\tau )$ manifold. In physical terms,  an internal
angular momentum has been generated by $N^{\nu}$ throughout the
space-time  in a reparameterization invariant way. This leads to a
{\it spin} in the fundamental point particle excitations,
resulting in a magnetic moment and subsequently the above
observations follow smoothly. Indeed, this is very appealing since
spin is a very fundamental and well studied geometrical property.
This enlargement is, in spirit, somewhat akin to the Superspace
formulation. Indeed, it would be interesting if a field theory can
be developed in this extended manifold and matched to the
conventional NC theory. Construction of the model and its
applications  constitute the first part of the paper.

In the second part, we comment on the established notion, (which
is used throughout in string theory context \cite{book}), that
{\it classically} the Nambu-Goto (square root action) and Polyakov
(induced metric action) forms are completely equivalent. We
explicitly demonstrate that in the presence of constraints,
showing the equivalence between the theories resulting from the
above two pictures raises subtle points and in fact the Polyakov
form happens to be more general than the Nambu-Goto form, in the
sense that the NC structure is fixed  in the Nambu-Goto form,
whereas it is gauge choice dependent in the Polyakov form. Lastly,
we make a brief observation on the quantum theory.

As mentioned before, our model consists of the coordinate
$X^{\mu}(\sigma ,\tau )$ coupled to another field $N^{\mu}(\sigma
,\tau )$ in a reparaterization invariant way. The $N^{\mu}$
manifold is compact with $N^{\mu}N_{\mu}=1$. Similar to the
extension of the relativistic point particle action to the string
action, our model can be thought of as an extension of the
relativistic {\it spinning} point particle \cite{j,ch} with the
spatial parameter $\sigma$ being infinite in extent as well.
$N^{\mu}$ imparts the spin.

Let us  consider the Lagrangian,
\begin{equation}
{\cal L}=2[(\dot X\dot N)(X'N')-(\dot XN')(X'\dot N)]^{\frac{1}{2}}\equiv 2{\cal A},
\label{l}
\end{equation}
where $(\dot XN')\equiv
\frac{dX^{\mu}}{d\tau}\frac{dX_{\mu}}{d\sigma}$ etc. and the
conjugate momenta are
\begin{equation}
P^{\mu}=\frac{\partial {\cal L}}{\partial\dot X_{\mu}}=\frac{1}{{\cal A}}[(X'N')\dot N^{\mu}-(X'\dot N)N'^{\mu}]~;~
P_{N}^{\mu}=\frac{\partial {\cal L}}{\partial\dot N_{\mu}}=\frac{1}{{\cal A}}[(X'N')\dot X^{\mu}-(\dot XN')X'^{\mu}],
\label{mom}
\end{equation}
The primary constraints in the theory, indicating reparamerization invariance, are
\begin{equation}
P^{\mu}X'_{\mu}\equiv 0~;~P_N^{\mu}N'_{\mu}\equiv 0~.
\label{cons}
\end{equation}
The resulting First Class Constraint (FCC) algebra
\footnote {In the Dirac Hamiltonian constraint analysis \cite{dirac} the commuting
(in the Poisson Bracket sense) constraints are termed as First Class Constraints (FCC) and the
non-commuting
ones as Second Class Constraints (SCC). The former signals gauge invariance and the latter
modify the symplectic structure from Poisson Brackets to Dirac Brackets.}
(or Schwinger algebra) is given by,
$$\chi_{1,2}=(PX')\pm (P_NN')$$
$$
\{\chi_1(\sigma ),\chi _1(\sigma ')\}=(\chi_1(\sigma )+\chi _1(\sigma '))\delta '(\sigma-\sigma')~;
~\{\chi_2(\sigma ),\chi _2(\sigma ')\}=(\chi_1(\sigma )+\chi _1(\sigma '))\delta '(\sigma-\sigma'),$$
\begin{equation}
\{\chi_1(\sigma ),\chi _2(\sigma ')\}=(\chi_2(\sigma )+\chi _2(\sigma '))\delta '(\sigma-\sigma'),
\label{sch}
\end{equation}
where the Poisson brackets are,
\begin{equation}
\{X^{\mu}(\sigma),P^{\nu}(\sigma')\}=g^{\mu\nu}\delta(\sigma-\sigma')~,~
\{N^{\mu}(\sigma),P_N^{\nu}(\sigma')\}=g^{\mu\nu}\delta(\sigma-\sigma').
\label{pb}
\end{equation}
In a geometric interpretation, just as the string action constitutes the elementary
area in the $\sigma - \tau $ plane, the action from (\ref{l}) can be worked out to represent
an area in the $\sigma - \tau $ plane (see appendix). So far, this is a free
theory with (decoupled) wave equations satisfied by $X^{\mu}$ and $N^{\mu}$. This will be
shown later.

The $N^{\mu}$ manifold is now compactified by invoking the
constraint $\chi_{3} \equiv N^{2}-1\approx 0$ via the multiplier
$\lambda$,
\begin{equation}
{\cal L}=2{\cal A}+\lambda (N^{2}-1).
\label{lc}
\end{equation}
As stated before, the model is motivated by an earlier spinning particle model \cite{ch},
which was later amended \cite{j} to a first order Hamiltonian form.  $\chi_{3}$ leads to  more
primary  constraints, the full set being,
\begin{equation}
\chi_{1}~;~\chi_{2}~;~\chi_{3}\equiv N^{2}-1 ~;~\psi_{1}\equiv (PN)~;~\psi_{2}\equiv (PP_N)-(X'N')~,
\label{con}
\end{equation}
out of which $\chi$s and $\psi$ are FCCs and SCCs \cite{dirac}
respectively. The canonical Hamiltonian vanishes on the constraint
surface, as it should in a reparameterization invariant theory.
The following definition of Dirac Bracket \cite{dirac},
\begin{equation}
\{A,B\}_{DB}=\{A,B\}-\{A,\psi_{i}\}\{\psi_{i},\psi_{j}\}^{-1}\{\psi_{j},B\};~~i,j=1,2~,
\label{db}
\end{equation}
yields the new symplectic structure \footnote {Unless otherwise stated, henceforth all $\{,\}$
are Dirac brackets.} ,
$$
\{X_{\mu}(\sigma),X_{\nu}(\sigma')\}=\frac{{P_N}_{\mu}N_{\nu}-{P_N}_{\nu}N_{\mu}}{P^{2}+N'^{2}}\delta ~;~\{X_{\mu}(\sigma),P_{\nu}(\sigma')\}=g_{\mu\nu}\delta -\frac{N_{\mu } N'_{\nu}}{P^{2}+N'^{2}}(\sigma)\delta' $$
\begin{equation}
\{N_{\mu}(\sigma),{P_N}_{\nu}(\sigma')\}=(g_{\mu\nu}-\frac{P_{\mu}P_{\nu}}{P^{2}+N'^{2}})\delta~;~\{{P_N}_{\mu}(\sigma),{P_N}_{\nu}(\sigma')\}=(\frac{P_{\mu} X'_{\nu}}{P^{2}+N'^{2}}(\sigma)+\frac{P_{\nu} X'_{\mu}}{P^{2}+N'^{2}}(\sigma'))\delta'
\label{db1}
\end{equation}

$$
\{X_{\mu}(\sigma),N_{\nu}(\sigma')\}=-\frac{N_{\mu}P_{\nu}}{P^{2}+N'^{2}}
\delta~;~\{X_{\mu}(\sigma),{P_N}_{\nu}(\sigma')\}=-\frac{{P_N}_{\mu}P_{\nu}}{P^{2}+N'^{2}}\delta -\frac{N_{\mu} X'_{\nu}}{P^{2}+N'^{2}}(\sigma)\delta'~, $$
\begin{equation}
\{{P_N}_{\nu}(\sigma),P_{\mu}(\sigma')\}=\frac{P_{\nu}N'_{\mu}}{P^{2}+N'^{2}}(\sigma)\delta'.
\label{db2}
\end{equation}
The notations in the above are $\delta ' \equiv \partial_{\sigma} \delta (\sigma -\sigma ')~,~\frac{A(\sigma )}{B(\sigma )}=\frac{A}{B}(\sigma )$. Using the above brackets it is straightforward to check that
\begin{equation}
J_{\mu\nu}=\int d\sigma(P_{\mu}X_{\nu}-P_{\nu}X_{\mu}+{P_N}_{\mu}N_{\nu}-{P_N}_{\nu}N_{\mu})
\label{j}
\end{equation}
generates the angular momentum algebra and transforms the vectors properly,
\begin{equation}
\{J_{\mu\nu},V_{\alpha}(\sigma )\}=g_{\nu\alpha}V_{\mu}(\sigma)-g_{\mu\alpha}V_{\nu}(\sigma),
\label{vec}
\end{equation}
where $V_{\alpha}\equiv
X_{\alpha},P_{\alpha},N_{\alpha},{P_N}_{\alpha}.$ Clearly the spin
contribution is coming from the $N^{\mu}$ sector. Thus we have
constructed the NC space-time $X^{\mu}$ as is evident from the
non-vanishing $\{X,X\}$ bracket. The NC factor is not a constant,
which would have violated Lorentz invariance. Introduction of spin
has turned the ordinary space-time into an NC one on the $\sigma -
\tau $ plane. Thus, we have succeeded in constructing a model for
the noncommutative plan and (\ref{db2}) is the cherished form of
the noncommutative structure. This is the major result of the
present work.

A quite involved computation reveals that the Schwinger algebra (\ref{sch}) is
intact if one exploits the new symplectic structure given in (\ref{db1},\ref{db2}).
This establishes the fact that the diffeomorphism symmetry is {\it not} destroyed by the
introduction of a constraint ($\chi_{3}$ in the present case) from outside. We will use
this idea in a crucial way later in the Polyakov formulation.

Now that the field theoretic model for the NC plane is at hand, we
can check whether it leads to some of the intriguing features
\cite{jab,big} of elementary (NC) excitations, in a fundamental
way. Note that although these observations \cite{jab,big} have
been made in connection with string theory, what really matters is
the underlying NC spacetime structure. In our more general field
theoretic formulation of the NC spacetime, these results are
reproduced naturally. To discuss the manifestations of
noncommutativity in our toy model, we concentrate on  the (so
called) point particles living on the NC plane constructed here to
analyze how this NC has affected them. However, deriving the
particle properties from the involved symplectic structure given
in (\ref{db1},\ref{db2}), this turns out to be non-trivial.

For a conventional field theory, with the canonical equal-time  Poisson bracket
\begin{equation}
\{Q (\sigma ),P(\sigma ')\}=\delta (\sigma -\sigma '),
\label{can}
\end{equation}
the particle-like properties are revealed upon Fourier mode expansion,
\begin{equation}
Q (\sigma ,\tau )=\frac{1}{\sqrt{L}}\sum_{k}q_{k}(\tau)e^{ik\sigma},
\label{c1}
\end{equation}
where for convenience we have confined the system inside $L$.
Using the identity for discrete $k$
\begin{equation}
\frac{1}{L}\int e^{i(k-k')\sigma}d\sigma =\delta_{k,k'},
\label{c2}
\end{equation}
the discrete mode $q_{k}$ and its conjugate momentum $p_{k}$ are  expressed as
\begin{equation}
q_{k}(\tau)=\frac{1}{\sqrt{L}}\int e^{-ik\sigma}Q (\sigma ,\tau)d\sigma,~~
p_{k}(\tau)=\frac{1}{\sqrt{L}}\int e^{ik\sigma}P (\sigma ,\tau)d\sigma,
\label{c3}
\end{equation}
it is easy to verify that
\begin{equation}
\{q_{k},p_{k'}\}=\delta_{k,k'}.
\label{c4}
\end{equation}
In the above, all the modes are decoupled and behave in an
identical canonical fashion. Notice that for the $k=0$ oscillator,
this result is obtained trivially. By utilizing the following
relations,
\begin{equation}
q_{0}(\tau)=\frac{1}{\sqrt{L}}\int Q (\sigma ,\tau)d\sigma,~~
p_{0}(\tau)=\frac{1}{\sqrt{L}}\int P (\sigma ,\tau)d\sigma,
\label{c5}
\end{equation}
we can derive,
\begin{equation}
\{q_{0},p_{0}\}=\frac{1}{L}\int d\sigma d\sigma '\{Q (\sigma ),P (\sigma ' )\}=
\frac{1}{L}\int d\sigma =1.
\label{c6}
\end{equation}
In the present case, performing a similar analysis for the $k=0$ modes of all the field
variables, we find that (\ref{db1},\ref{db2}) reduces to
 a much simpler algebra,
$$
\{x_{\mu},x_{\nu}\}=\frac{{p_n}_{\mu}n_{\nu}-{p_n}_{\nu}n_{\mu}}{p^{2}} ~;~\{x_{\mu},p_{\nu}\}=g_{\mu\nu}~;~
\{n_{\mu},{p_n}_{\nu}\}=(g_{\mu\nu}-\frac{p_{\mu}p_{\nu}}{p^{2}}),$$
\begin{equation}
\{x_{\mu},n_{\nu}\}=-\frac{n_{\mu}p_{\nu}}{p^{2}}~;~\{x_{\mu},{p_n}_{\nu}\}=-\frac{{p_n}_{\mu}p_{\nu}}{p^{2}},
\label{db3}
\end{equation}
where $x^{\mu}(\tau)\equiv x^{\mu}_0(\tau)=\frac{1}{L}\int d\sigma
X_{\mu }(\sigma ,\tau)$ etc..
Naively this is obtained from (\ref{db1},\ref{db2}) by integrating
out $\sigma$, which amounts to dropping the $\sigma$-derivative
terms (since $k=0$) and $\delta '$. Clearly the brackets for the
non-zero $k$-modes are more complicated.

Notice that the above set (\ref{db3}) is identical to the Dirac Brackets given in the
corrected version of the spinning particle model \cite{j}. This
algebra can be thought of to be originated from the first order
Lagrangian posited in \cite{j} with $p^2$ denoting the mass of the
particle. In this sense, our model in 2+1-dimensions can be
interpreted as a {\it{field theoretic model for the anyon}},
excitations having arbitrary spin and statistics \cite{j,ch}.
Similar type of spin induced NC in a variant \cite{sg1} of the
present model has been discussed in \cite{sg2}.

The $\{x,x\}$-noncommutativity that appears in (\ref{db3}) is
actually a generalization of the more restricted form that is
commonly used, where $\{x_{\mu},x_{\nu}\}=\theta_{\mu\nu}$, $\theta_{\mu\nu}$ being a constant  c-number tensor. However,
non-constant and operatorial noncommutativity, of a different
form, have also appeared in \cite{dop}. On the other hand, as we
discuss below, reducing the noncommutativity in our model to the
above constant form is quite subtle.

Since we have already related our brackets (\ref{db1},\ref{db2})
to that of the spinning particle model symplectic structure
\cite{j}, let us rely more heavily on \cite{j} where a first order
Hamiltonian formalism has been discussed. Leaving out the details,
(which are provided in \cite{j}), the equations of motion are
obtained as,
\begin{equation}
\dot p_{\mu}=0~,~\dot n_{\mu}=0~\rightarrow p_{\mu}(\tau)\equiv p^0_{\mu},~
n_{\mu}(\tau)\equiv n^{0}_{\mu}.
\label{r2}
\end{equation}
One further gets,
\begin{equation}
\dot x_{\mu}=2\Lambda p^0_{\mu}~,~\dot {p_{n}}_{\mu}=-\lambda n^0_{\mu},
\label{r3}
\end{equation}
with $\Lambda $ and $\lambda$ being arbitrary. The equations
(\ref{r3}) are integrated to
\begin{equation}
x_{\mu}(\tau )=\alpha (\tau)p^0_{\mu}+x^0_{\mu}~,~{p_{n}}_{\mu}(\tau)=\beta (\tau)n^0_{\mu}+{p_n}^{0}_\mu ,
\label{r4}
\end{equation}
where
$\dot \alpha =2\Lambda ~,~\dot \beta =-\lambda $.
Exploiting the relation (\ref{r4}) in the noncommutativity bracket, we get
$$
\{x_{\mu},x_{\nu}\}=\frac{{p_n}_{\mu}n_{\nu}-{p_n}_{\nu}n_{\mu}}{p^{2}}=
\frac{1}{p^{2}}[(\beta n^0_{\mu}+{p_n}^0_\mu )n^0_\nu-(\beta n^0_{\nu}+{p_n}^0_\nu )n^0_{\mu}]$$
\begin{equation}
=\frac{{p_n}^0_{\mu}n^0_{\nu}-{p_n}^0_{\nu}n^0_{\mu}}{p^2}.
\label{r5}
\end{equation}
Clearly the right hand side is time independent. Finally, using
(\ref{r4}), we derive the complete symplectic structure that is
stable under time ($\tau$) translation,
$$
\{x^{0}_{\mu},x^{0}_{\nu}\}=\frac{{p_n}^{0}_{\mu}n^{0}_{\nu}-{p_n}^{0}_{\nu}n^{0}_{\mu}}{p^{2}} ~;~\{x^{0}_{\mu},p^{0}_{\nu}\}=g_{\mu\nu}~;~
\{n^{0}_{\mu},{p_n}^{0}_{\nu}\}=(g_{\mu\nu}-\frac{p^{0}_{\mu}p^{0}_{\nu}}{p^{2}}),$$
\begin{equation}
\{x^{0}_{\mu},n^{0}_{\nu}\}=-\frac{n^{0}_{\mu}p^{0}_{\nu}}{p^{2}}~;~\{x^{0}_{\mu},{p_n}^{0}_{\nu}\}=-\frac{{p_n}^{0}_{\mu}p^{0}_{\nu}}{p^{2}}.
\label{db5}
\end{equation}

The constant c-number NC parameter
$\{x_{\mu},x_{\nu}\}=\theta_{\mu\nu}$ is finally generated by
considering the spontaneous symmetry breaking reasoning,
originally developed in the context of Standard Model extension
\cite{cos1}, and later used in demonstrating the violation of
Lorentz invariance in  NC field theory \cite{car}. Here, the
vector fields attain a non-vanishing expectation value in the
vacuum at low energy.{\footnote{In the special case of the target
space being $2+1$-dimensional  \cite{j,ch}, the spin sector can be
removed altogether by exploiting the constraints. This leads to
the NC algebra $\{x^{\mu},x^{\nu}\}\approx
\epsilon^{\mu\nu\lambda}p_{\lambda}.$  In particular, this means
that $\rightarrow \{x^{1},x^{2}\}\approx p_0\approx m\equiv
constant$. Hence, this is compatible with the constant c-number
$x_{\mu}$-noncommutativity, at least in the non-relativistic
limit. It is not clear whether similar phenomenon will occur in
higher dimensions.}}

Let us now discuss the effects of the noncommutativity exhibited
in (\ref{db5}) in the quantum theory. However, (\ref{db5}) is not
convenient for the conventional quantization programme of
elevating the classical brackets to quantum commutators via the
correspondence principle. To facilitate this, we move on to a
canonical $(q^{\mu},Q^{\nu})$ setup with
 $$[q,q]=[Q,Q]=0~;~[q^\mu ,Q^{\nu}]=ig^{\mu\nu}$$ and solve the non-trivial spacetime
 algebra in (\ref{db5})
 by rewriting \cite{cnp},
 \begin{equation}
 x^{\mu}=q^{\mu}-\frac{1}{2}\theta^{\mu\nu}Q_{\nu}~,~p^{\mu}=Q^{\mu}.
 \label{q}
 \end{equation}
Thus the original theory should be reexpressed in the $q,Q$ variables in the quantized version.
This momentum dependent shift, a hallmark of NC quantum theory \cite{big}, appears here as a
prerequisite for quantization.

Returning to more down to earth physics, the 3+1-dimensional Coulomb potential due to a point
charge $e$ in the noncommutative spacetime now turns out to be
\begin{equation}
\frac{e}{\mid x_i\mid}=\frac{e}{[(q_i^2-\theta^{ij}q_iQ_j+O(\theta ^2)]^{\frac{1}{2}}}\approx \frac{e}{\mid q_i\mid}(1-\frac{\theta^{ij}q_iQ_j}{2q_i^2}+O(\theta ^2)).
\label{di}
\end{equation}
For simplicity, we have considered $\theta^{0i}=0$ in the above relation. Clearly the second
term in (\ref{di}) reflects the dipole nature of the excitation, with a
dipole moment $d_i=-\frac{1}{2}e\theta_{ij}Q^j$ \cite{jab}.

The dipole feature is also apparent if we place the point charge
$e$ in an external electrostatic potential $\phi$. The energy $W$
of the system in the NC plane is
\begin{equation}
W= e\phi (x_i)= e\phi (q_i-\frac{1}{2}\theta_{ij}Q^j) = e\phi (q)-\frac{1}{2}e\theta_{ij}Q^j\partial ^i\phi (q) +O(\theta^{2})= e\phi (q)-d_iE^i +O(\theta^{2}),
\label{ed}
\end{equation}
where $E^i=-\partial^{i}\phi$ is the electric field.
The dipole energy term is again reproduced. This constitutes the first part of the work.

Obviously the Polyakov formulation of a reparameterization
invariant theory is more transparent than the Nambu-Goto version
involving a square root action, that we have studied so far. In
this part we discuss the Polyakov formulation of the above model
and ascertain how far it is justified to consider the (previous)
Nambu-Goto and Polyakov forms as equivalent even in a classical
scenario. (This means we are not considering the quantum
anomalies.) The Polyakov form of the unconstrained model (\ref{l})
is,
\begin{equation}
{\cal L_{P}}=-\sqrt{-\gamma}\gamma^{ab}\partial_{a}X_\mu\partial_{b}N^{\mu}.
\label{pol}
\end{equation}
The non-dynamical metric $\gamma^{ab}$ can be eliminated to
reproduce the Nambu-Goto Lagrangian (\ref{l}) in the conventional
way. Diffeomorphism and Weyl invariances allow us to enforce locally the
conformal gauge, $\gamma^{ab}=diag (-1,1)$, which reduces
(\ref{pol}) to a simple form,
\begin{equation}
{\cal L_P}=-[-(\dot X\dot N)+(X'N')].
\label{conf}
\end{equation}
This will lead to two decoupled free wave equations for $X^\mu $ and $N^\mu $, as mentioned
before. However the constraint $\chi_{3}$ will change the dynamics of $X^\mu$  to,
\begin{equation}
\ddot X^\mu -X''^\mu-N_\nu (\ddot X^\nu -X''^\nu )N^\mu =0.
\label{new}
\end{equation}
The Hamiltonian is
\begin{equation}
{\cal H_P}=(PP_N)+(X'N')~;~P_{\mu}=\dot N_{\mu}~,~{P_N}_{\mu}=\dot X_{\mu}.
\label{h}
\end{equation}
Since we are considering a classical theory, for the moment we
ignore the fact that the Hamiltonian in (\ref{h}) is not positive
definite and invoke the constraint $\chi_{3}=N^2-1$ on this model.
Note that exploiting the conformal gauge {\it before} invoking the
constraint $\chi_{3}$ is justified from our previous experience of
working in Nambu-Goto picture where we saw that $\chi_{3}$ does
not spoil the invariances. The conformal gauge removes the reparameterization
invariance and we must check the stability of $\chi_{3}$ under time translation.
However, time persistence of the
successive constraints now generates an infinite chain of
constraints of the following form,
$$
\chi_{3}=N^2-1\rightarrow (NP)\rightarrow P^2-N'^2 \rightarrow (PN'')-(P'N') \rightarrow P'^2-N''^2 \rightarrow ... $$
\begin{equation}
... \rightarrow (P'N'')-(P''N'') \rightarrow P''^2-N'''^2 \rightarrow ...~~.
\label{fcc}
\end{equation}
Notice that {\it all} the constraints in (\ref{fcc}) are in involution, {\it i.e.} the
constraints are FCC, the phase space being canonical (\ref{pb}). Quite clearly this (Polyakov)
constrained system is very different from the finite number of FCC and SCCs present in the
Nambu-Goto version given in (\ref{con}). In fact one can choose a gauge in the Nambu-Goto
form \cite{hrt} which is equivalent to the conformal gauge in the Polyakov form. However,
the structure of the SCCs (inducing the NC) is fixed in the Nambu-Goto form.
 This is precisely the disparity, even in the classical scenario, between  Nambu-Goto and
 Polyakov formulations in the presence of constraints, that we had set out to establish.

A correspondence between the two formulations can be obtained if  in the Polyakov form we choose a suitable gauge. In particular let us choose $\psi_{2}=(PP_N)-(X'N')$ of (\ref{con}) as the gauge condition. In the chain (\ref{fcc}), it will keep the FC nature of $\chi_{3}=N^2-1$ intact. The gauge condition $\psi_{2}$ together with $(NP)=\psi_{1}$ are rendered to an  SCC pair. Subsequently rest of the FCCs will drop out of the picture once the Dirac brackets are exploited, which are precisely the ones derived before (\ref{db1},\ref{db2}).
Hence, we recover the constraint structure of the Nambu-Goto form (\ref{con}) without the
FCCs $\chi_{1}$ and $\chi_{2}$, since we have fixed the conformal gauge. Thus, in the Polyakov
form, whether NC is induced or not, depends on our choice of gauge. Obviously any
$P^\mu $-independent gauge will fail to generate NC in $X^\mu$. In this sense, we conclude
that the Polyakov form is more general than the Nambu-Goto form.

Lastly, let us comment on the non-positive definite nature of the Hamiltonian in (\ref{h})
which can hinder its quantization. A possible way out is to invoke the ideas of t'Hooft \cite{h} where we rewrite $H_P=\int \cal {H_P}$ in the following way,
\begin{equation}
H_P=H_+-H_-~,~\{H_+,H_-\}=0,
\label{hoof}
\end{equation}
where both $H_\pm$ are positive definite. Subsequently one quantizes the system by
taking $H_+$ as the Hamiltonian and employ $H_-$ as a constraint such
that $H_-\mid Physical ~ State >=0$. Thus we can express ${\cal {H_P}}$ in (\ref{h}) as,
\begin{equation}
{\cal {H_P}}=\frac{1}{4}[((P+P_N)^2+(X'+N')^2)-((P-P_N)^2+(X'-N')^2)]\equiv {\cal {H_+}}-{\cal {H_-}},
\label{pm}
\end{equation}
and proceed with the constraint analysis of ${\cal {H_P}}$.

Finally, let us make a passing comment on the yet to be investigated quantum theory. Note
that the classical conformal invariance of the model will be broken by quantum anomalies.
However, since the model is new, the structure of the anomaly and also whether
any non-trivial anomaly vanishing constraints emerge, are some of the topics of interest.

To conclude, we have constructed a  {\it{field theoretic}} toy
model, which yields a noncommutative space-time, without any
external influence. The noncommutativity is induced by the
constraints of the theory via Dirac brackets. We have also derived
the symplectic structure for the zero momentum Fourier modes,
which coincides with the algebra of a spinning particle model
\cite{j}. The algebra for the non-zero momentum sector is much
more complicated.

To put our work in the proper perspective, it should be stressed
that generation of the spinning particle model \cite{j} is
actually a by product of our construction and not the main issue
involved. What we have achieved here is an explicit field
theoretic construction of a noncommutative spacetime. This can
serve as an alternative to the noncommutative spacetime where
{\it{commutative}} ({\it{i.e.}} ordinary) spacetime coordinates
can be used at the expense of working in an extended phase space
\cite{bt} with extra auxiliary degrees of freedom. The
noncommutativity was induced via the Dirac brackets \cite{dirac},
the latter being necessary since the system has Second Class
constraints \cite{dirac}. In the extended space \cite{bt}, these
constraints are modified to First Class constraints \cite{dirac}
and so the Dirac brackets are not needed. Incidentally, the
noncanonical (and especially operator valued) Dirac brackets are a
hindrance to the subsequent quantization of the model. It is very
important to note that the extended space \cite{bt} is (by
construction) completely canonical, which a prerequisite for
carrying out the canonical quantization programme. The
noncommutativity is reproduced through the auxiliary variables.
The extended space formulation of the present model will proceed
along the lines of \cite{sg2,bt}. Although straightforward, this
is a separate problem by itself and is postponed for a future
publication.

In this model, various traits of a string induced noncommutative
quantum theory on the $D$-brane, such as dipolar nature of the
basic excitations \cite{jab} and momentum dependent shifts in the
interaction point \cite{big} appear naturally in our
noncommutative spacetime. The observation \cite{sw} that
noncommutative and ordinary field theories are alternative
descriptions of the same underlying theory, is corroborated here
by showing that they are gauge equivalent. This has been achieved
by the introduction of an additional spin field which endows the
point charges with a spin and subsequently a magnetic moment.

Also, treating the present model as an explicit example, we show
that, in the presence of constraints,  exact equivalence between
the Nambu-Goto and Polyakov formulations can not be established in
a naive way even classically.

{\bf {Appendix:}} The target space consists of "position" vectors
$x^{i}\equiv \{X^{\mu},N^{\nu}\}$ with the metric,
$$g^{ij}=
 \left (
\begin{array}{cc}
 0^{\mu}_{\nu}  &  \delta^{\mu}_{\nu}\\
\delta^{\mu}_{\nu} &  0^{\mu}_{\nu}
\end{array}
\right )
$$
where $0^{\mu}_{\nu}$ and $\delta ^{\mu}_{\nu}$ represent  null
and unit matrices respectively. Hence, the distance $ds$ in
spacetime $(ds)^{2}=g_{ij}dx^{i}dx^{j}=dX^{\mu}dN_{\mu}$ leads to
the induced metric,
$$(ds)^{2}=\gamma_{ab}d\xi^{a}d\xi^{b}~~;~~\gamma_{ab}=\frac{\partial X^{\mu}}{\partial \xi^{a}}\frac{\partial N_{\mu}}{\partial \xi^{b}}~;~~\xi^{a}\equiv (\tau ,\sigma ).$$
Thus $\sqrt{-det ~ \gamma}$ reproduces the action (\ref{l}).
Furthermore, restricting the $x^{i}$-space such that the induced
metric $\gamma_{ab}$ is symmetric, one obtains the interpretation
of $\sqrt{-det ~ \gamma} ~d\tau d\sigma $ as the infinitesimal
area in the $\tau - \sigma$ plane trivially \cite{hat}.

The off-diagonal nature of the target space metric requires some
comments. The motivation was the construction of a
reparameterization invariant geometric field theory to describe a
noncommutative spacetime and the structure of the metric is geared
for that purpose. It might be useful to note that prior to the introduction of the compactifying constraint on $N_\mu$, the $X_\mu$ and $N_\mu$ sectors are treated in an identical way, which is reflected by their symmetrical appearence in the expressions. Unfortunately, at the present level of research
concerning this model, it is difficult to ascribe a physical
significance regarding the form of the metric.

\vskip .5cm {\bf Acknowledgement:} It is a pleasure to thank
Professor Roman Jackiw for his suggestion that the first order
formulation of the spinning particle model should be exploited.
Also, I am grateful to the referee for the pertinent questions raised.

\end{document}